\documentclass[prd,showkeys,floatfix,twocolumn,amsmath,amssymb,floatfix]{revtex4}
\usepackage{graphicx,color,dcolumn,booktabs,bm}
\usepackage{longtable,lscape}
\usepackage{amssymb}
\usepackage{indentfirst}
\usepackage{subfigure}
\usepackage{epsfig}
\usepackage{feynmf}
\usepackage{epstopdf}
\usepackage{slashed}
\usepackage{cases}
\usepackage{multirow}
\usepackage[colorlinks,citecolor=blue,anchorcolor=red,menucolor=red,linkcolor=red,filecolor=red,runcolor=red,urlcolor=blue,frenchlinks=red]{hyperref}

\begin{document}
%=====================================================================================
%=====================================================================================
\title{Strong decays of the $\phi(2170)$ as a fully-strange tetraquark state}
%=====================================================================================
%=====================================================================================
%

\author{Yi-Wei Jiang$^1$}
\author{Wei-Han Tan$^1$}
\author{Hua-Xing Chen$^1$}
\email{hxchen@seu.edu.cn}
\author{Er-Liang Cui$^2$}
\email{erliang.cui@nwafu.edu.cn}

\affiliation{
$^1$School of Physics, Southeast University, Nanjing 210094, China
\\
$^2$College of Science, Northwest A\&F University, Yangling 712100, China
}

\begin{abstract}
We study strong decays of the $\phi(2170)$, along with its possible partner $X(2436)$, as two fully-strange tetraquark states of $J^{PC} = 1^{--}$. We consider seven decay channels: $\phi \eta$, $\phi \eta^\prime$, $\phi f_0(980)$, $\phi f_1(1420)$, $h_1(1415) \eta$, $h_1(1415) \eta^\prime$, and $h_1(1415) f_1(1420)$. Some of these channels are kinematically possible, and we calculate their relative branching ratios through the Fierz rearrangement. Future experimental measurements on these ratios can be useful in determining the nature of the $\phi(2170)$ and $X(2436)$. The $\phi(2170)$ has been observed in the $\phi f_0(980)$, $\phi \eta$, and $\phi \eta^\prime$ channels, and we propose to further examine it in the $h_1(1415) \eta$ channel. Evidences of the $X(2436)$ have been observed in the $\phi f_0(980)$ channel, and we propose to verify whether this structure exists or not in the $\phi \eta$, $\phi \eta^\prime$, $h_1(1415) \eta$, and $h_1(1415) \eta^\prime$ channels.
\end{abstract}
\pacs{12.39.Mk, 12.38.Lg, 12.40.Yx}
\keywords{fully-strange tetraquark, Fierz rearrangement}
\maketitle
\pagenumbering{arabic}

\section{Introduction}
\label{sec:intro}

In the traditional quark model we can categorize hadrons into $\bar q q$ mesons and $qqq$ baryons~\cite{pdg}. In recent years many exotic hadrons were observed in particle experiments, which can not be easily explained in the traditional quark model~\cite{Chen:2016qju,Liu:2019zoy,Chen:2022asf,Lebed:2016hpi,Esposito:2016noz,Hosaka:2016pey,Guo:2017jvc,Ali:2017jda,Olsen:2017bmm,Karliner:2017qhf,Bass:2018xmz,Brambilla:2019esw,Guo:2019twa,Ketzer:2019wmd,Yang:2020atz,Fang:2021wes,Jin:2021vct,JPAC:2021rxu,Meng:2022ozq,Brambilla:2022ura}, such as the charmonium-like states $X(3872)$ of $I^GJ^{PC} = 0^+1^{++}$~\cite{Belle:2003nnu}, $Y(4220)$ of $I^GJ^{PC} = 0^-1^{--}$~\cite{BaBar:2005hhc,BESIII:2016bnd}, and $Z_c(3900)$ of $I^GJ^{PC} = 1^+1^{+-}$~\cite{BESIII:2013ouc,Belle:2013yex}. However, there are not so many exotic hadrons in the light sector that only contain the $up/down/strange$ quarks. The $\phi(2170)$ of $I^GJ^{PC} = 0^-1^{--}$, also denoted as $Y(2175)$, is one of them. It is often taken as the strangeonium counterpart of the $Y(4220)$ owing to their similarities in production mechanism and decay patterns.

The $\phi(2170)$ was first observed in 2006 by the BaBar Collaboration via the initial state radiation process $e^+ e^-\rightarrow \gamma_{\rm{ISR}}\phi f_0(980)$~\cite{BaBar:2006gsq,BaBar:2007ptr,BaBar:2007ceh,BaBar:2011btv}. Later it was confirmed by Belle in the $e^+ e^-\rightarrow \phi \pi^+ \pi^-$ and $e^+ e^-\rightarrow \phi f_0(980)$ processes~\cite{Belle:2008kuo}, and it was also observed by BESII/BESIII in the $J/\psi \rightarrow \eta \phi f_0(980)$ process~\cite{BES:2007sqy,BESIII:2014ybv,BESIII:2017qkh}. According to the latest version of PDG~\cite{pdg}, its mass and width were averaged to be:
\begin{eqnarray}
\phi(2170)/Y(2175) &:& M = 2163 \pm 7 {\rm~MeV} \, ,
\\ \nonumber      && \Gamma = 103_{-21}^{+28}{\rm~MeV} \, .
\end{eqnarray}
In recent years various experimental studies on the $\phi(2170)$ were carried out by the BESIII Collaboration in the direct $e^+ e^-$ annihilation to the $\phi \eta$, $\phi \eta^\prime$, $\phi K^+ K^-$, $K^+ K^-$, $\omega \eta$, $K^0_S K^0_L$, and $\phi\pi^+\pi^-$ final states~\cite{BESIII:2021bjn,BESIII:2020gnc,BESIII:2019ebn,BESIII:2018ldc,BESIII:2020xmw,BESIII:2021yam,BESIII:2023fmx}, etc. In Ref.~\cite{BESIII:2020vtu} a partial wave analysis of the $e^+ e^- \rightarrow K^+ K^- \pi^0 \pi^0$ process was performed by BESIII, indicating that the $\phi(2170)$ has a sizable partial width to $K^+(1460)K^-$,$K^+_1(1400)K^-$, and $K^+_1(1270)K^-$, but a much smaller partial width to $K^{*+}(892)K^{*-}(892)$ and $K^{*+}(1410)K^-$.

Since its discovery, the $\phi(2170)$ has stimulated many theoretical methods and models to explain its nature. Possible interpretations of this interesting structure are abundant and diverse, including the traditional $s\bar s$ meson as an exited state~\cite{Ding:2007pc,Wang:2012wa,Afonin:2014nya,Pang:2019ttv,Zhao:2019syt,Li:2020xzs}, a strangeonium hybrid state~\cite{Ding:2006ya,Ho:2019org}, a fully-strange tetraquark state~\cite{Wang:2006ri,Chen:2008ej,Drenska:2008gr,Deng:2010zzd,Chen:2018kuu,Ke:2018evd,Agaev:2019coa,Liu:2020lpw}, a hidden-strangeness baryon-antibaryon state strongly coupling to the $\Lambda\bar\Lambda$ channel~\cite{Abud:2009rk}, a bound state of $\Lambda\bar\Lambda$~\cite{Klempt:2007cp,Zhao:2013ffn,Deng:2013aca,Dong:2017rmg,Yang:2019mzq}, and a dynamically generated state in the $\phi K K$ and $\phi \pi \pi$ systems~\cite{Napsuciale:2007wp,Gomez-Avila:2007pgn,MartinezTorres:2008gy} or in the $\phi f_0(980)$ system~\cite{Alvarez-Ruso:2009vkn,Coito:2009na}. Within the lattice QCD formalism, the authors of Ref.~\cite{Dudek:2011bn} studied the $\phi(2170)$ under the hybrid hypothesis, but their results do not favor this interpretation. Besides, productions of the $\phi(2170)$ were studied in Refs.~\cite{Bystritskiy:2007wq,Ali:2011qi} by using the Nambu-Jona-Lasinio model and the Drell-Yan mechanism, and its decay properties were studied in Refs.~\cite{Chen:2011cj,Chen:2023jaj,Malabarba:2020grf,Malabarba:2023zez,Li:2020xzs,Feng:2021igh} by using the initial single pion emission mechanism, the dispersion theory, the three-hadron interactions, and the $^3 P_0$ model.

In addition, the $\phi(2170)$ may have a partner state at around 2.4~GeV, denoted as $X(2436)$. Its evidences have been observed in the $\phi f_0(980)$ and $\phi\pi^+\pi^-$ channels by the BaBar, Belle, BESII, and BESIII experiments~\cite{BaBar:2007ptr,BES:2007sqy,Belle:2008kuo,BESIII:2014ybv}. The authors of Ref.~\cite{Shen:2009mr} performed a combined fit to the data of BaBar and Belle, where the mass and width of this structure were measured to be
\begin{eqnarray}
X(2436) &:& M = 2436 \pm 34 {\rm~MeV} \, ,
\\ \nonumber      && \Gamma = 99 \pm 105{\rm~MeV} \, ,
\end{eqnarray}
when fitting the $\phi f_0(980)$ cross section, but its statistical significance is less than $3 \sigma$. Recently, the BESIII Collaboration further studied this structure through the $e^+ e^- \rightarrow\phi \pi^+ \pi^-$ process~\cite{BESIII:2021lho}, but its statistical significance is no more than $2\sigma$. Therefore, more experimental studies are necessary to clarify whether the $X(2436)$ exists or not.

Although there are considerable efforts from both the experimental and theoretical sides, the nature of the $\phi(2170)$ and $X(2436)$ is still not clear. In order to clarify their nature, it is useful to examine their decay modes and relative branching ratios. Especially, it is useful to study the $\phi(2170)$ decays into the $\phi\eta$ and $\phi\eta^\prime$ channels, in order to investigate the ratio
\begin{equation}
R_{\eta/\eta^\prime}^{\rm{exp}} \equiv {\mathcal{B}_{\phi\eta}^{Y}\Gamma_{e^+e^-}^{Y} \over \mathcal{B}_{\phi\eta^\prime}^{Y}\Gamma_{e^+e^-}^{Y}} \, ,
\end{equation}
where
\begin{eqnarray}
\mathcal{B}_{\phi\eta}^{Y} &\equiv& {\rm Br}( \phi(2170) \to \phi\eta) \, ,
\\
\mathcal{B}_{\phi\eta^\prime}^{Y} &\equiv& {\rm Br}( \phi(2170) \to \phi\eta^\prime) \, ,
\\
\Gamma_{e^+e^-}^{Y} &\equiv& \Gamma( e^+e^- \to \phi(2170)) \, .
\end{eqnarray}
In Refs.~\cite{BESIII:2020gnc,BESIII:2021bjn} the BESIII Collaboration separately studied the $e^+e^- \to \phi\eta/\phi\eta^\prime$ processes and extracted:
\begin{eqnarray}
\mathcal{B}_{\phi\eta}^{Y}\Gamma_{e^+e^-}^{Y} &=& \left\{\renewcommand{\arraystretch}{1.3}\begin{array}{l}0.24_{-0.07}^{+0.12}~{\rm{eV}}~~~~~\text{(sol I)},\\ 10.11_{-3.13}^{+3.87}~{\rm{eV}}~~~\text{(sol II)},\end{array}\right.
\label{BES1}
\\
\mathcal{B}_{\phi\eta^\prime}^{Y}\Gamma_{e^+e^-}^{Y} &=& 7.1 \pm 0.7 \pm 0.7~{\rm{eV}}~~\text{(sol I/II)},
\label{BES2}
\end{eqnarray}
where ``sol I/II'' denote the two possible solutions. The $e^+e^- \to \phi\eta$ process has also been investigated by BaBar~\cite{BaBar:2007ceh}:
\begin{equation}
\mathcal{B}_{\phi\eta}^{Y}\Gamma_{e^+e^-}^{Y} = 1.7 \pm 0.7 \pm 1.3~{\rm{eV}},
\end{equation}
and Belle~\cite{Belle:2022fhh}:
\begin{equation}
\mathcal{B}_{\phi\eta}^{Y}\Gamma_{e^+e^-}^{Y} = \left\{\renewcommand{\arraystretch}{1.3}\begin{array}{l} 0.09 \pm 0.05~{\rm{eV}}~~~\text{(sol I)},\\ 0.06 \pm 0.02~{\rm{eV}}~~~\text{(sol II)}, \\ 16.7 \pm 1.2~{\rm{eV}}~~~~~\text{(sol III)},\\ 17.0 \pm 1.2~{\rm{eV}}~~~~~\text{(sol IV)}.\end{array}\right.
\end{equation}
Based on Eqs.~(\ref{BES1}) and (\ref{BES2}), we can derive
\begin{eqnarray}
{\rm{BESIII:}}~~R_{\eta/\eta^\prime}^{\rm{exp}} &=& \left\{\renewcommand{\arraystretch}{1.3}\begin{array}{l}0.034_{-0.014}^{+0.029}~~~~\text{(sol I)},\\ 1.42_{-0.60}^{+1.03}~~~~~~~\text{(sol II)}.\end{array}\right.
\label{BES}
\end{eqnarray}
Theoretically, this ratio was calculated in Ref.~\cite{Malabarba:2023zez} to be $R_{\eta/\eta^\prime}= 2.6\sim5.2$, where the $\phi(2170)$ was considered as a dynamically-generated state from the $\phi f_0(980)$ interaction. More theoretical calculations on this ratio are helpful to reveal the nature of the $\phi(2170)$.

We have applied the method of QCD sum rules to study the $\phi(2170)$ and $X(2436)$ in Refs.~\cite{Chen:2008ej,Chen:2018kuu}. In Ref.~\cite{Chen:2008ej} we systematically constructed the fully-strange tetraquark currents and found only two independent ones. We separately used them to perform QCD sum rule analyses by calculating only the diagonal two-point correlation functions. In Ref.~\cite{Chen:2018kuu} we further calculated the off-diagonal two-point correlation functions, and the obtained results can explain both the $\phi(2170)$ and $X(2436)$ as two fully-strange tetraquark states. In this paper we shall utilize the Fierz rearrangement method~\cite{Chen:2019wjd,Chen:2019eeq,Chen:2020pac} to study their strong decays as the fully-strange tetraquark states of $J^{PC} = 1^{--}$.

This paper is organized as follows. In Sec.~\ref{sec:current} we construct the fully-strange tetraquark currents of $J^{PC} = 1^{--}$ within the diquark-antidiquark picture. We use them to further construct two mixing currents that are non-correlated, which can be used to simultaneously interpret the $\phi(2170)$ and $X(2436)$ as two fully-strange tetraquark states. We apply the Fierz rearrangement to transform these two mixing currents into the meson-meson currents, based on which we study the decay behaviors of the $\phi(2170)$ and $X(2436)$ in Sec.~\ref{sec:decay}. The obtained results are discussed and summarized in Sec.~\ref{sec:summary}.

\section{Currents and Fierz Identities}
\label{sec:current}

\begin{figure}[hbt]
\begin{center}
\subfigure[]{\includegraphics[width=0.15\textwidth]{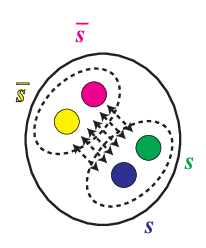}}
~~~~~~~~~~
\subfigure[]{\includegraphics[width=0.15\textwidth]{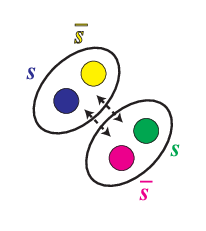}}
\caption{Two types of fully-strange tetraquark currents: (a) the diquark-antidiquark currents $\eta(x,y)$ and (b) the meson-meson currents $\xi(x,y)$.}
\label{fig:current}
\end{center}
\end{figure}

The fully-strange tetraquark currents with the quantum number $J^{PC} = 1^{--}$ have been systematically constructed and studied in Refs.~\cite{Chen:2008ej,Chen:2018kuu}, where we consider two types of tetraquark currents, as illustrated in Fig.~\ref{fig:current}:
\begin{eqnarray}
\eta(x,y)   &=& [s^T_a(x) \mathbb{C} \Gamma_1 s_b(x)] \times [\bar s_c(y) \Gamma_2 \mathbb{C} \bar s_d^T(y)] \, ,
\\ \xi(x,y) &=& [\bar s_a(x) \Gamma_3 s_b(x)]         \times [\bar s_c(y) \Gamma_4 s_d(y)] \, .
\end{eqnarray}
Here $\Gamma_i$ are Dirac matrices, the subscripts $a \cdots d$ are color indices, $\mathbb{C} = i\gamma_2 \gamma_0$ is the charge-conjugation operator, and the superscript $T$ represents the transpose of Dirac indices. We call the former $\eta(x,y)$ diquark-antidiquark currents and the latter $\xi(x,y)$ meson-meson currents, which will be separately investigated in the following subsections.

\subsection{Diquark-antidiquark currents and their mixing}
\label{sec:mixing}

There are two fully-strange diquark-antidiquark interpolating currents with the quantum number $J^{PC} = 1^{--}$:
\begin{eqnarray}
%-------------------------------------eta 1------------------------------------
&& \eta_{1\mu} =
\\ \nonumber && ~~~ (s_a^T \mathbb{C} \gamma_5 s_b) (\bar{s}_a \gamma_\mu \gamma_5 \mathbb{C} \bar{s}_b^T) - (s_a^T \mathbb{C} \gamma_\mu \gamma_5 s_b) (\bar{s}_a \gamma_5 \mathbb{C} \bar{s}_b^T) \, ,
\label{def:eta1}
%-------------------------------------eta 2------------------------------------
\\ && \eta_{2\mu} =
\\ \nonumber && ~~~ (s_a^T \mathbb{C} \gamma^\nu s_b) (\bar{s}_a \sigma_{\mu\nu} \mathbb{C} \bar{s}_b^T) - (s_a^T \mathbb{C} \sigma_{\mu\nu} s_b) (\bar{s}_a \gamma^\nu \mathbb{C} \bar{s}_b^T) \, .
\label{def:eta2}
\end{eqnarray}
These two currents are independent of each other.

In Ref.~\cite{Chen:2008ej} we separately use $\eta_{1\mu}$ and $\eta_{2\mu}$ to perform QCD sum rule analyses, where we calculate only the diagonal correlation functions:
\begin{equation}
\langle 0 | \eta_{1\mu} { \eta_{1\nu}^\dagger } | 0 \rangle~~~{\rm and }~~~\langle 0 | \eta_{2\mu} { \eta_{2\nu}^\dagger } | 0 \rangle \, .
\end{equation}
However, in Ref.~\cite{Chen:2018kuu} we find that the off-diagonal correlation function
\begin{equation}
\langle 0 | \eta_{1\mu} { \eta_{2\nu}^\dagger } | 0 \rangle \neq 0 \, ,
\end{equation}
is also non-zero, indicating that $\eta_{1\mu}$ and $\eta_{2\mu}$ are correlated with each other, so they can couple to the same physical state. To deal with this, in Ref.~\cite{Chen:2018kuu} we further construct two mixing currents:
\begin{eqnarray}
J_{1\mu} &=& \cos\theta~\eta_{1\mu} + \sin\theta~i~\eta_{2\mu} \, ,
\label{def:J1}
\\ J_{2\mu} &=& \sin\theta~\eta_{1\mu} + \cos\theta~i~\eta_{2\mu} \, .
\label{def:J2}
\end{eqnarray}
When setting the mixing angle to be $\theta = -5.0^{\rm o}$, these two currents satisfy
\begin{equation}
\langle 0 | J_{1\mu} { J_{2\nu}^\dagger} | 0 \rangle
\left\{
\begin{array}{c}
\ll \langle 0 | J_{1\mu} J_{1\nu}^\dagger | 0 \rangle
\\ \ll \langle 0 | J_{2\mu} J_{2\nu}^\dagger | 0 \rangle
\end{array}
\right.\, ,
\end{equation}
with the threshold value around $s_0 \approx 6.0$ GeV$^2$ and the Borel mass around $M_B^2 \approx 2.5$ GeV$^2$. This condition indicates that the two currents $J_{1\mu}$ and $J_{2\mu}$ are non-correlated, {\it i.e.}, they can not mainly couple to the same state $Y$, otherwise,
\begin{eqnarray}
\nonumber \langle 0 | J_{1\mu} J_{2\nu}^\dagger | 0 \rangle &\equiv& \sum_n \delta(s-M^2_n) \langle 0| J_{1\mu} | n \rangle \langle n | J_{2\nu}^\dagger |0 \rangle + \cdots
\\ \nonumber &\approx& \delta(s-M^2_Y) \langle 0| J_{1\mu} | Y \rangle \langle Y | J_{2\nu}^\dagger |0 \rangle + \cdots
\\ &\neq& 0 \, .
\end{eqnarray}
Accordingly, we assume that $J_{1\mu}$ and $J_{2\mu}$ mainly couple to two different states $Y_1$ and $Y_2$ through
\begin{eqnarray}
\langle 0| J_{1\mu} | Y_1 \rangle &=& f_{Y_1}~\epsilon_{\mu} \, ,
\label{eq:coupling1}
\\ \langle 0| J_{2\mu} | Y_2 \rangle &=& f_{Y_2}~\epsilon_{\mu} \, ,
\label{eq:coupling2}
\end{eqnarray}
where $f_{Y_1}$ and $f_{Y_1}$ are the decay constants, and $\epsilon_{\mu}$ is the polarization vector.

In Ref.~\cite{Chen:2018kuu} we use $J_{1\mu}$ and $J_{2\mu}$ to perform QCD sum rule analyses. When setting the working regions to be 5.0~GeV$^2< s_0 < 7.0$~GeV$^2$ and 2.0~GeV$^2 < M_B^2 < 4.0$~GeV$^2$, we calculate the masses of $Y_1$ and $Y_2$ to be
\begin{eqnarray}
M_{Y_1} &=& 2.41 \pm 0.25 {\rm~GeV} \, ,
\\ \nonumber
M_{Y_2} &=& 2.34 \pm 0.17 {\rm~GeV} \, ,
\end{eqnarray}
with the mass splitting
\begin{equation}
\Delta M = 71 ^{+172}_{-~48}  {\rm~MeV} \, .
\end{equation}

The mass extracted from $J_{2\mu}$ is consistent with the experimental mass of the $\phi(2170)$, indicating its possible explanation as the fully-strange tetraquark state $Y_2$. The QCD sum rule result extracted from the non-correlated current $J_{1\mu}$ suggests that the $\phi(2170)$ may have a partner state whose mass is about $2.41 \pm 0.25$~GeV. This value is consistent with the experimental mass of the $X(2436)$, indicating its possible explanation as the fully-strange tetraquark state $Y_1$. We shall further study their strong decays through the two mixing currents $J_{1\mu}$ and $J_{2\mu}$ in Sec.~\ref{sec:decay}.

\subsection{Meson-meson currents and Fierz rearrangement}
\label{sec:fierz}

Besides the diquark-antidiquark currents $\eta_{1\mu}$ and $\eta_{2\mu}$, we can also construct the fully-strange meson-meson currents. There are four fully-strange meson-meson interpolating currents with the quantum number $J^{PC} = 1^{--}$:
\begin{eqnarray}
\xi_{1\mu} &=& (\bar{s}_a s_a)(\bar{s}_b \gamma_\mu s_b) \, ,
\label{def:xi1}
\\ \xi_{2\mu} &=& (\bar{s}_a \gamma^\nu\gamma_5 s_a)(\bar{s}_b \sigma_{\mu\nu}\gamma_5 s_b) \, ,
\label{def:xi2}
\\ \xi_{3\mu} &=& {\lambda_{ab}}{\lambda_{cd}}(\bar{s}_a s_b)(\bar{s}_c \gamma_\mu s_d) \, ,
\label{def:xi3}
\\ \xi_{4\mu} &=& {\lambda_{ab}}{\lambda_{cd}} (\bar{s}_a \gamma^\nu\gamma_5 s_b)(\bar{s}_c \sigma_{\mu\nu}\gamma_5 s_d) \, .
\label{def:xi4}
\end{eqnarray}
We can derive through the Fierz rearrangement that only two of them are independent, {\it e.g.},
%
%%%%%%%%%%%%%%%%%%%%%%%%%%%%%%%%%%%%%%%%%%%%%%%%%%%%%%%%%%%%%%%%%%%%%%%%%%%%%%
\begin{eqnarray}
\xi_{3\mu} &=& - \frac{5}{3} \xi_{1\mu} - i \xi_{2\mu} \, , \, \,
\\ \nonumber
\xi_{4\mu} &=& 3i \xi_{1\mu} + \frac{1}{3} \xi_{2\mu} \, .
\end{eqnarray}
%%%%%%%%%%%%%%%%%%%%%%%%%%%%%%%%%%%%%%%%%%%%%%%%%%%%%%%%%%%%%%%%%%%%%%%%%%%%%%
%
Moreover, we can also derive through the Fierz rearrangement the relations between the diquark-antidiquark currents $\eta_i$ and the meson-meson currents $\xi_i$:
%
%%%%%%%%%%%%%%%%%%%%%%%%%%%%%%%%%%%%%%%%%%%%%%%%%%%%%%%%%%%%%%%%%%%%%%%%%%%%%%
\begin{eqnarray}
\eta_{1\mu} &=& - \xi_{1\mu} + i \xi_{2\mu} \, ,
\label{eq:fierz1}
\\
\eta_{2\mu} &=& 3i \xi_{1\mu} - \xi_{2\mu} \, .
\label{eq:fierz2}
\end{eqnarray}
%%%%%%%%%%%%%%%%%%%%%%%%%%%%%%%%%%%%%%%%%%%%%%%%%%%%%%%%%%%%%%%%%%%%%%%%%%%%%%
%
Therefore, these two constructions are equivalent with each other, but note that this equivalence is just between the local diquark-antidiquark and meson-meson currents, while the tightly-bound diquark-antidiquark tetraquark states and the weakly-bound meson-meson molecular states are significantly different. To well describe them, we need the non-local currents, but we are still not able to use them to perform QCD sum rule analyses yet.

We can use Eqs.~(\ref{eq:fierz1}) and (\ref{eq:fierz2}) to transform the mixing currents $J_{1\mu}$ and $J_{2\mu}$ to be
\begin{eqnarray}
J_{1\mu} &=& - 0.74 \xi_{1\mu} + 1.08 i \xi_{2\mu} \, ,
\label{eq:fierz3}
\\
J_{2\mu} &=&  - 2.90 \xi_{1\mu} - 1.08 i \xi_{2\mu} \, .
\label{eq:fierz4}
\end{eqnarray}
These two Fierz identities will be used in Sec.~\ref{sec:decay} to study the strong decays of the two states $Y_1$ and $Y_2$.

\subsection{Strangeonium operators and decay constants}
\label{sec:constant}

\begin{table*}[hbt]
\begin{center}
\renewcommand{\arraystretch}{2}
\caption{Couplings of the strangeonium operators to the strangeonium states. Color indices are omitted for simplicity.}
\begin{tabular}{c | c | c | c | c | c}
\hline\hline
~~~Operators~~~ & ~~~$J^{PC}$~~~ & ~~~Mesons~~~ & ~~~$J^{PC}$~~~ & ~~~~~~~~~~~~~~Couplings~~~~~~~~~~~~~~ & ~~~~~~~~~~~Decay Constants~~~~~~~~~~~
\\ \hline\hline
$J^S = \bar s s$              & $0^+0^{++}$         & $f_0(980)$                   & $0^+0^{++}$         & $\langle 0 | J^S | f_0 \rangle = m_{f_0} f_{f_0}$
                                                                                                         & $f_{f_0} = 358$~MeV~\mbox{\cite{Cheng:2019tgh}}
\\ \hline
\multirow{2}{*}{$J^P = \bar s i\gamma_5 s$}
                              & \multirow{2}{*}{$0^+0^{-+}$}   & $\eta$            & $0^+0^{-+}$         & $\langle0| J^P | \eta \rangle = \lambda_{\eta}$
                                                                                                         & $\lambda_{\eta} \approx 218$~MeV$^2$
\\ \cline{3-6}
                              &                                & $\eta^\prime$     & $0^+0^{-+}$         & $\langle0| J^P | \eta^\prime \rangle = \lambda_{\eta^\prime}$
                                                                                                         & $\lambda_{\eta^\prime} \approx 638$~MeV$^2$
\\ \hline
$J^V_\mu=\bar s\gamma_\mu s$  & $0^-1^{--}$         & $\phi$                       & $0^-1^{--}$         & $\langle 0 | J^V_\mu | \phi \rangle = m_\phi f_\phi \epsilon_\mu$
                                                                                                         & $f_{\phi} \approx 233$~MeV~\mbox{\cite{Bijnens:1996nq,Becirevic:2003pn}}
\\ \hline
\multirow{3}{*}{$J^A_\mu = \bar s \gamma_\mu \gamma_5 s$}
                              & \multirow{3}{*}{$0^+1^{++}$} & $\eta$              & $0^+0^{-+}$         & $\langle0| J^A_\mu | \eta \rangle = i q_\mu f_{\eta}$
                                                                                                         & $f_{\eta} \approx 159$~MeV
\\ \cline{3-6}
                              &                              & $\eta^\prime$       & $0^+0^{-+}$         & $\langle0| J^A_\mu | \eta^\prime \rangle = i q_\mu f_{\eta^\prime}$
                                                                                                         & $f_{\eta^\prime} \approx 200$~MeV
\\ \cline{3-6}
                              &                              & $f_1(1420)$         & $0^+1^{++}$         & $\langle0| J^A_\mu |f_1 \rangle = f_{f_1} m_{f_1} \epsilon_\mu$
                                                                                                         & $f_{f_1} = 217$~MeV~\mbox{\cite{Yang:2007zt}}
\\ \hline
\multirow{2}{*}{$J^T_{\mu\nu} = \bar s \sigma_{\mu\nu} s$}
                              & \multirow{2}{*}{$0^-1^{\pm-}$} & $\phi$            & $0^-1^{--}$         & $\langle 0 | J^T_{\mu\nu} | \phi \rangle = i f^T_{\phi} (p_\mu\epsilon_\nu - p_\nu\epsilon_\mu) $
                                                                                                         & $f_{\phi}^T \approx 175$~MeV~\mbox{\cite{RBC-UKQCD:2008mhs}}
\\ \cline{3-6}
                              &                                & $h_1(1415)$       & $0^-1^{+-}$         & $\langle 0 | J^T_{\mu\nu} | h_1 \rangle = i f^T_{h_1} \epsilon_{\mu\nu\alpha\beta} \epsilon^\alpha p^\beta $
                                                                                                         & $f_{h_1}^T = f_{b_1}^T \times {f_\phi \over f_{\rho}} = 194$~MeV~\mbox{\cite{Ball:1996tb}}
\\ \hline\hline
\end{tabular}
\label{tab:coupling}
\end{center}
\end{table*}

The meson-meson currents $\xi_1$ and $\xi_2$ are both composed of two strangeonium operators, whose couplings to the strangeonium states have been studied in the literature to some extent~\cite{pdg,Cheng:2019tgh,Bijnens:1996nq,Becirevic:2003pn,Yang:2007zt,RBC-UKQCD:2008mhs,Ball:1996tb}, as summarized in Table~\ref{tab:coupling}. Especially, we follow Refs.~\cite{Leutwyler:1997yr,Kaiser:1998ds,Escribano:2005qq,Escribano:2015nra,Escribano:2015yup,Schechter:1992iz,Kiselev:1992ms,Herrera-Siklody:1997pgy,Bass:2018xmz,Bali:2021qem} to study the axial-vector operator $J^A_\mu = \bar s \gamma_\mu \gamma_5 s$, and use the two-angle mixing formalism to describe the pseudoscalar mesons $\eta$ and $\eta^\prime$ as
\begin{eqnarray}
|\eta\rangle &=& \cos\theta_8 |\eta_8\rangle - \sin \theta_0 | \eta_0 \rangle + \cdots \, ,
\\ \nonumber
|\eta^\prime\rangle &=& \sin\theta_8 |\eta_8\rangle + \cos \theta_0 | \eta_0 \rangle + \cdots \, ,
\end{eqnarray}
where
\begin{eqnarray}
|\eta_8 \rangle &=& | u \bar u + d \bar d - 2 s \bar s \rangle/\sqrt6 \, ,
\\ \nonumber
|\eta_0 \rangle &=& | u \bar u + d \bar d + s \bar s \rangle/\sqrt3 \, ,
\end{eqnarray}
and $\cdots$ denotes the other components such as the pseudoscalar glueball and charmonium, etc.

We define the flavor octet and singlet axial-vector operators as
\begin{eqnarray}
A_\mu^{8} &=& \left( { \bar u \gamma_\mu \gamma_5 u +  \bar d \gamma_\mu \gamma_5 d - 2 \bar s \gamma_\mu \gamma_5 s }\right)/\sqrt{12} \, ,
\\ \nonumber
A_\mu^{0} &=& \left( { \bar u \gamma_\mu \gamma_5 u +  \bar d \gamma_\mu \gamma_5 d + \bar s \gamma_\mu \gamma_5 s }\right)/\sqrt6 \, ,
\end{eqnarray}
which couple to $\eta$ and $\eta^\prime$ through
\begin{equation}
\langle0| A_\mu^a | P(q) \rangle = i q_\mu f_P^a \, .
\end{equation}
Here $f_P^a$ is the matrix for the decay constants
\begin{equation}
\left(\begin{array}{cc}
f_\eta^8 & f_\eta^0
\\
f_{\eta^\prime}^8 & f_{\eta^\prime}^0
\end{array}\right)
=
\left(\begin{array}{cc}
f_8 \cos\theta_8 & - f_0 \sin\theta_0
\\
f_8 \sin\theta_8 &   f_0 \cos\theta_0
\end{array}\right) \, ,
\end{equation}
where~\cite{Ali:1997ex,Feldmann:1997vc}
\begin{eqnarray}
\nonumber \theta_8 &=& - 22.2^\circ \, ,
\\ \theta_0 &=& - 9.1^\circ \, ,
\label{parameter1}
\\ \nonumber f_8 &=& 168 {\rm~MeV} \, ,
\\ \nonumber f_0 &=& 157 {\rm~MeV} \, .
\end{eqnarray}

Based on the above formula, we can derive
\begin{eqnarray}
\langle0| J^A_\mu | \eta(q) \rangle &=& i q_\mu f_{\eta} \, ,
\\ \nonumber
\langle0| J^A_\mu | \eta^\prime(q) \rangle &=& i q_\mu f_{\eta^\prime} \, ,
\end{eqnarray}
where
\begin{eqnarray}
f_{\eta} &\approx& 159 {\rm~MeV}\, ,
\\ \nonumber
f_{\eta^\prime} &\approx& 200 {\rm~MeV} \, .
\end{eqnarray}

We can further approximate the couplings of the pseudoscalar operator $J^P = \bar s i\gamma_5 s$ to $\eta$ and $\eta^\prime$ as
\begin{eqnarray}
\langle0| J^P | \eta(q) \rangle &=& \lambda_{\eta} \, ,
\\ \nonumber
\langle0| J^P | \eta^\prime(q) \rangle &=& \lambda_{\eta^\prime} \, ,
\end{eqnarray}
where
\begin{eqnarray}
\lambda_{\eta} &\approx& {6 f_{\eta} m_{\eta} \over m_u + m_d + 4 m_s} = 218 {\rm~MeV} \, ,
\\ \nonumber
\lambda_{\eta^\prime} &\approx& {3 f_{\eta^\prime} m_{\eta^\prime} \over m_u + m_d + m_s} = 638 {\rm~MeV} \, .
\end{eqnarray}

\section{Relative branching ratios}
\label{sec:decay}

\begin{figure}[hbt]
\begin{center}
\includegraphics[width=0.25\textwidth]{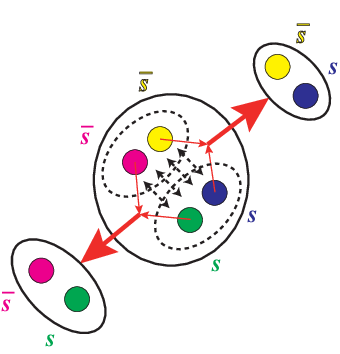}
\caption{The fall-apart decay process of a fully-strange tetraquark state to two strangeonium mesons.}
\label{fig:diagram}
\end{center}
\end{figure}

In this section we study the strong decays of the fully-strange tetraquark states with the quantum number $J^{PC} = 1^{--}$. As depicted in Fig.~\ref{fig:diagram}, when one quark meets one antiquark and the other quark meets the other antiquark at the same time, a fully-strange tetraquark state can fall apart to two strangeonium mesons. This process can be described by the Fierz identities given in Eqs.~(\ref{eq:fierz3}) and (\ref{eq:fierz4}).

Let us start with Eq.~(\ref{eq:fierz3}) and perform analyses qualitatively. The strangeonium operators $J^S = \bar{s}_a s_a$ and $J^V_\mu = \bar{s}_b \gamma_\mu s_b$ couple to the $f_0(980)$ and $\phi(1020)$, respectively. Hence, the meson-meson current $\xi_{1\mu}$ well couples to the $\phi f_0(980)$ channel, and the mixing current $J_{1\mu}$ also couples to this channel. Accordingly, the state $Y_1$ can decay to this channel. Similarly, we can derive six other possible channels to be $\phi \eta$, $\phi \eta^\prime$, $\phi f_1(1420)$, $h_1(1415) \eta$, $h_1(1415) \eta^\prime$, and $h_1(1415) f_1(1420)$. Among them, the $\phi \eta$, $\phi \eta^\prime$, $h_1(1415) \eta$, and $h_1(1415) \eta^\prime$ channels are kinematically allowed.

In principle, we need the coupling of $J_{1\mu}$ to $Y_1$ as an input to quantitatively calculate the partial decay widths of these channels. This parameter has been defined in Eq.~(\ref{eq:coupling1}) as $f_{Y_1}$. However, it is not necessary if we just want to calculate the relative branching ratios. We still take Eq.~(\ref{eq:fierz3}) as an example, from which we can extract the couplings of the mixing current $J_{1\mu}$ to the $\phi f_0(980)$ and $\phi \eta$ channels:
\begin{eqnarray}
&& \langle 0 | J_{1\mu} | \phi(p_1,\epsilon_1)~f_0(p_2) \rangle
\\ \nonumber &=& -0.74 \times \epsilon_1^\mu ~ m_{f_0} f_{f_0} ~ m_\phi f_\phi \, ,
\\
&& \langle 0 | J_{1\mu} | \phi(p_1,\epsilon_1)~\eta(p_2) \rangle
\\ \nonumber &=& 0.54 \times f_{\eta} f^T_{\phi} \epsilon_{\mu\nu\alpha\beta}p_2^\nu( p_1^\alpha \epsilon_1^\beta - p_1^\beta \epsilon_1^\alpha) \, .
\end{eqnarray}
Then we can extract the couplings of the state $Y_1$ to the $\phi f_0(980)$ and $\phi \eta$ channels:
\begin{eqnarray}
&& \langle Y_1(p,\epsilon) | \phi(p_1,\epsilon_1)~f_0(p_2) \rangle
\\ \nonumber &=& -0.74 c \times \epsilon \cdot \epsilon_1 ~ m_{f_0} f_{f_0} ~ m_\phi f_\phi \, ,
\\
&& \langle Y_1(p,\epsilon) | \phi(p_1,\epsilon_1)~\eta(p_2) \rangle
\\ \nonumber &=& 0.54 c \times f_{\eta} f^T_{\phi} \epsilon_{\mu\nu\alpha\beta} \epsilon^\mu p_2^\nu ( p_1^\alpha \epsilon_1^\beta - p_1^\beta \epsilon_1^\alpha) \, .
\end{eqnarray}
The overall factor $c$ is related to the decay constant $f_{Y_1}$. After calculating the partial decay widths $\Gamma_{Y_1 \to \phi f_0(980)}$ and $\Gamma_{Y_1 \to \phi \eta}$, we can eliminate this factor and obtain
\begin{equation}
{\mathcal{B}(Y_1 \to \phi f_0(980)) \over \mathcal{B}(Y_1 \to \phi \eta)} = 1.14 \, .
\end{equation}
Similarly, we can investigate the $\phi \eta^\prime$, $h_1(1415) \eta$, and $h_1(1415) \eta^\prime$ channels to obtain:
\begin{eqnarray}
\nonumber \mathcal{B}(\,Y_1 &\to& ~\phi \eta ~:~ \phi \eta^\prime ~:~ \phi f_0 ~: h_1(1415) \eta : h_1(1415) \eta^\prime\,)
\\ &=& 1.00 : \,0.71\, :\,1.14\,: ~~~~0.74~~~~ :~0.32  \, .
\label{eq:Y1}
\end{eqnarray}
The above calculations are done within the naive factorization scheme, so our uncertainty is significantly larger than the well-developed QCD factorization scheme~\cite{Beneke:1999br,Beneke:2000ry,Beneke:2001ev}. However, our calculations are done after eliminating the ambiguous overall factor $f_{Y_1}$, which largely reduces our uncertainty.

\begin{figure*}[htb]
\begin{center}
\scalebox{1.06}{\includegraphics{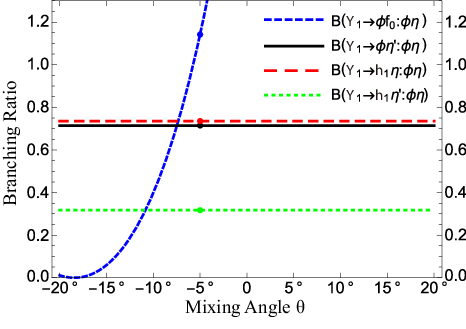}}
\label{fig:ratiosy1}
~~~
\scalebox{0.85}{\includegraphics{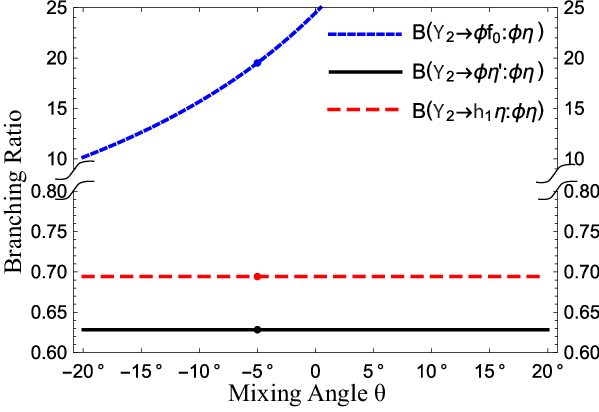}}
\label{fig:ratiosy2}
\end{center}
\caption{Relative branching ratios of the fully-strange tetraquark states $Y_1$ (left) and $Y_2$ (right) with respect to the mixing angle $\theta$, with the $\phi \eta$, $\phi \eta^{\prime}$, $\phi f_0(980)$, $h_1(1415) \eta$, and $h_1(1415) \eta^\prime$ channels taken into account.
\label{fig:ratios}}
\end{figure*}

It is interesting to examine the dependence of the above ratios on the mixing angle $\theta$, as shown in the left panel of Fig.~\ref{fig:ratios}. Especially, the ratio
\begin{equation}
R_{\eta/\eta^\prime}^{Y_1} \equiv {\mathcal{B}(Y_1 \to \phi \eta) \over \mathcal{B}(Y_1 \to \phi \eta^\prime)} = 1.40 \, ,
\end{equation}
does not depend on this parameter. This ratio can be useful in clarifying the nature of the $X(2436)$ as a fully-strange tetraquark state.

Following the same procedures, we study the strong decays of the state $Y_2$ through the mixing current $J_{2\mu}$. In this case we consider the $\phi f_0(980)$, $\phi \eta$, $\phi \eta^\prime$, and $h_1(1415) \eta$ channels, since the $h_1(1415) \eta^\prime$ channel is kinematically forbidden. Their relative branching ratios are calculated to be
\begin{eqnarray}
\nonumber \mathcal{B}(\,Y_2 &\to& ~~\phi \eta ~~:~~ \phi \eta^\prime ~~:~~ \phi f_0 ~~:~~ h_1(1415) \eta\,)
\\ &=& ~1.00~ : ~\,0.63~\, : ~19.52~ : ~~~~~~0.69  \, .
\label{eq:Y2}
\end{eqnarray}
We show the dependence of these ratios on the mixing angle $\theta$ in the right panel of Fig.~\ref{fig:ratios}. Again, the ratio
\begin{equation}
R_{\eta/\eta^\prime}^{Y_2} \equiv {\mathcal{B}(Y_2 \to \phi \eta) \over \mathcal{B}(Y_2 \to \phi \eta^\prime)} = 1.59 \, ,
\end{equation}
does not depend on the mixing angle $\theta$, and moreover, it is almost the same as the ratio $R_{\eta/\eta^\prime}^{Y_1} = 1.40$. This ratio can be useful in clarifying the nature of the $\phi(2170)$ as a fully-strange tetraquark state.

%================================================================================
%================================================================================
\section{Summary and discussions}
\label{sec:summary}
%================================================================================
%================================================================================

In this paper we systematically study the strong decays of the $\phi(2170)$ and $X(2436)$ as two fully-strange tetraquark states with the quantum number $J^{PC} = 1^{--}$. Their corresponding fully-strange tetraquark currents have been systematically constructed in our previous studies~\cite{Chen:2008ej,Chen:2018kuu}, where we consider both the diquark-antidiquark and meson-meson constructions. We have also derived their relations there through the Fierz rearrangement, which are used in the present study to study their strong decay properties.

There are two independent diquark-antidiquark currents, defined in Eqs.~(\ref{def:eta1}) and (\ref{def:eta2}) as $\eta_{1\mu}$ and $\eta_{2\mu}$. In Ref.~\cite{Chen:2008ej} we calculate their diagonal correlation functions, and in Ref.~\cite{Chen:2018kuu} we further calculate their off-diagonal correlation function. Based on the obtained results, we construct two mixing currents, defined in Eqs.~(\ref{def:J1}) and (\ref{def:J2}) as $J_{1\mu}$ and $J_{2\mu}$ with the mixing angle $\theta = -5.0^{\rm o}$. These two mixing currents are non-correlated with each other, so they separately couple to two different states $Y_1$ and $Y_2$, whose masses are calculated in Ref.~\cite{Chen:2018kuu} through the QCD sum rule method to be
\begin{eqnarray*}
M_{Y_1} &=& 2.41 \pm 0.25  {\rm~GeV} \, ,
\\ M_{Y_2} &=& 2.34 \pm 0.17  {\rm~GeV} \, .
\end{eqnarray*}
These two values are consistent with the experimental masses of the $X(2436)$ and $\phi(2170)$, indicating their possible explanations as the fully-strange tetraquark states $Y_1$ and $Y_2$, respectively. Accordingly, we can use the mixing currents $J_{1\mu}$ and $J_{2\mu}$ to further study their decay properties.

We use the Fierz rearrangement to transform the mixing currents $J_{1\mu}$ and $J_{2\mu}$ to be the combinations of the meson-meson currents $\xi_{1\mu}$ and $\xi_{2\mu}$, as defined in Eqs.~(\ref{def:xi1}) and (\ref{def:xi2}). The obtained Fierz identities are given in Eqs.~(\ref{eq:fierz3}) and (\ref{eq:fierz4}). Based on these results, we study the decay mechanism depicted in Fig.~\ref{fig:diagram}, where a fully-strange tetraquark state fall-apart decays to two strangeonium mesons. We consider altogether seven possible channels: $\phi \eta$, $\phi \eta^\prime$, $\phi f_0(980)$, $\phi f_1(1420)$, $h_1(1415) \eta$, $h_1(1415) \eta^\prime$, and $h_1(1415) f_1(1420)$. Some of these channels are kinematically possible, whose relative branching ratios are calculated to be:
\begin{eqnarray}
\nonumber \mathcal{B}(\,X(2436) &\to& \phi \eta : \phi \eta^\prime : \phi f_0 : h_1(1415) \eta : h_1(1415) \eta^\prime\,)
\\ &=& 1\, : \,0.71\, :\,1.14\,: ~~~0.74~~~ :0.32    \, .
\\
\nonumber \mathcal{B}(\,\phi(2170) &\to& \phi \eta : \phi \eta^\prime : \phi f_0 : h_1(1415) \eta\,)
\\ &=& 1\, : \,0.63\, : 19.52 : ~~~0.69  \, .
\end{eqnarray}
The $X(2436)$ and $\phi(2170)$ satisfy that
\begin{eqnarray}
R_{\eta/\eta^\prime}^{X(2436)} &\equiv& {\mathcal{B}(X(2436) \to \phi \eta) \over \mathcal{B}(X(2436) \to \phi \eta^\prime)} = 1.40 \, ,
\\
R_{\eta/\eta^\prime}^{\phi(2170)} &\equiv& {\mathcal{B}(\phi(2170) \to \phi \eta) \over \mathcal{B}(\phi(2170) \to \phi \eta^\prime)} = 1.59 \, ,
\end{eqnarray}
and moreover, these two ratios do not depend on the mixing angle $\theta$. Compared to the BESIII measurement listed in Eq.~(\ref{BES}), our theoretical results within the fully-strange tetraquark picture are consistent with their second solution $R_{\eta/\eta^\prime}^{\rm{exp}} = 1.42_{-0.60}^{+1.03}$.

The $\phi(2170)$ has been observed in the $\phi f_0(980)$, $\phi \eta$, and $\phi \eta^\prime$ channels. Our results suggest that it can also be searched for in the $h_1(1415) \eta$ channel. There are some evidences of the $X(2436)$ in the $\phi f_0(980)$ channel. Our results suggest that it can also be searched for in the $\phi \eta$, $\phi \eta^\prime$, $h_1(1415) \eta$, and $h_1(1415) \eta^\prime$ channels. We propose to examine these decay channels in the future Belle-II, BESIII, COMPASS, GlueX, J-PARC, and PANDA experiments. Especially, the ratios $R_{\eta/\eta^\prime}^{\phi(2170)} \approx R_{\eta/\eta^\prime}^{X(2436)} \approx 1.5$ can be useful in clarifying the nature of the $\phi(2170)$ and $X(2436)$ as the fully-strange tetraquark states with the quantum number $J^{PC} = 1^{--}$.

%
%=====================================================================================
%=====================================================================================
%=====================================================================================
\section*{Acknowledgments}
%=====================================================================================
%=====================================================================================
%=====================================================================================
%

This project is supported by
the National Natural Science Foundation of China under Grant No.~12075019 and No.~12005172,
the Jiangsu Provincial Double-Innovation Program under Grant No.~JSSCRC2021488,
and
the Fundamental Research Funds for the Central Universities.

\bibliographystyle{elsarticle-num}
\bibliography{ref}

\end{document}